\documentclass[conference]{IEEEtran}

\usepackage{cite}
\usepackage{amsmath, amssymb}
\usepackage{graphicx}
\usepackage{xcolor}
\usepackage{booktabs}
\usepackage{multirow}
\usepackage{url}
\usepackage[hidelinks]{hyperref}
\usepackage{siunitx}
\usepackage{makecell}

\usepackage{pgfplots}
\pgfplotsset{compat=1.18}

\RequirePackage{orcidlink}

\begin{document}

\title{When GPUs Fail Quietly:\\
Observability-Aware Early Warning Beyond Numeric Telemetry}

\author{
  \IEEEauthorblockN{%
    Michael Bidollahkhani\orcidlink{0000-0001-8122-4441},
    Freja Nordsiek\orcidlink{0000-0001-5185-3634},
    Julian M. Kunkel\,\orcidlink{0000-0002-6915-1179}}
  \IEEEauthorblockA{%
    Gesellschaft für wissenschaftliche Datenverarbeitung mbH Göttingen (GWDG)\\
    Georg-August-Universität Göttingen\\
    Göttingen, Germany\\
    e-mail: {michael.bkhani@uni-goettingen.de, freja.nordsiek@gwdg.de,
    julian.kunkel@gwdg.de}
} }

\maketitle

\begin{abstract}
GPU nodes are central to modern HPC and AI workloads, yet many operational failures do not manifest as immediate hard faults. While some instabilities emerge gradually as weak thermal or efficiency drift, a significant class occurs abruptly with little or no numeric precursor. In these detachment-class failures, GPUs become unavailable at the driver or interconnect level and the dominant observable signal is structural, including disappearance of device metrics and degradation of monitoring payload integrity.
This paper proposes an observability-aware early-warning framework that jointly models (i) utilization-aware thermal-drift signatures in GPU telemetry and (ii) monitoring-pipeline degradation indicators such as scrape latency increase, sample loss, time-series gaps, and device-metric disappearance. The framework is evaluated on production telemetry from GPU nodes at the Gesellschaft für wissenschaftliche Datenverarbeitung mbH Göttingen (GWDG), where GPU, node, monitoring, and scheduler logs can be correlated.
We anchor the analysis with an operator-curated incident catalog (Jan 2025–Feb 2026) and a GPU-class forensic pass (69 matched, 15 processed). The study focuses on seven GPU detachment incidents (“fallen off bus”), of which five have complete telemetry archives for alignment. Using reproducible slice definitions, fixed windowing, and budgeted anomaly detection under unlabeled conditions, we evaluate statistical and learning-based detectors via alert-budget metrics, weak-event lead time, and cross-node generalization.
Results show that detachment failures exhibit minimal numeric precursor and are primarily observable through structural telemetry collapse, while joint observability-aware modeling increases early-warning lead time compared to GPU-only detection. The outcome is a practical methodology and benchmark artifact for studying early-stage GPU reliability in production environments without precise timestamped component-level failure labels. Instead, the dataset provides an operator-curated incident catalog containing coarse failure categories and day-level timestamps, which are used to anchor analysis windows around operational events.
\end{abstract}


\section{Introduction}
GPU-node instability remains a persistent operational issue in HPC systems, particularly under sustained high load and mixed workloads. Common symptoms include thermal throttling, power capping, clock oscillations, VRAM stress, and intermittent driver or interconnect issues. While some instabilities develop gradually and produce weak numeric signals that can be observed before conventional alarms trigger, a significant class of GPU failures occurs abruptly, without pronounced precursor anomalies in standard telemetry.

In practice, abrupt GPU failures often manifest as device detachments at the driver or interconnect level (e.g., GPUs ``falling off the bus''). In such cases, temperatures, power consumption, clocks, and utilization may remain nominal until the failure event, after which device-level telemetry disappears partially or entirely. These detachment-class failures challenge value-based monitoring and anomaly detection approaches, as the dominant signal is not metric deviation but the sudden loss of expected metrics and scrape payload integrity.

A second operational challenge is observability degradation. Monitoring pipelines may degrade before or during instability, leading to elevated scrape durations, intermittent scrape failures, reduced telemetry sample counts, truncated metric sets, and time-series gaps. For detachment-class failures in particular, observability degradation is not merely a side effect but often the primary observable manifestation of the fault, yielding a coupled failure mode in which the system becomes unstable and simultaneously becomes hard or impossible to observe.

Operationally, GPU detachment events occur alongside other node-level incidents (kernel panic/softlock, watchdog resets, network/IB issues, hangs/offline resets, and memory/ECC/MCE events). We use an operator-curated incident list to define analysis windows around these events and then focus this paper’s forensic case studies on detachment-class GPU incidents, where observability collapse is often the dominant signal.

\subsection{Contributions}
This work makes four contributions:
\begin{itemize}
    \item A reproducible pipeline specification for extracting GPU telemetry, node signals, and monitoring indicators from large-scale HPC telemetry using a slice definition that is explicit and reviewable.
    \item Drift- and trend-based instability signatures with robustness constraints for low utilization and per-GPU baseline normalization.
    \item Observability degradation indicators that quantify monitoring quality loss and measurement gaps, and a multi-archive availability matrix to ensure valid plane comparisons.
    \item A replicable evaluation protocol for unlabeled production telemetry using alert budgets, weak-event lead time, and cross-node comparability.
\end{itemize}

%%%%%%%%%% RELATED WORKS HERE %%%%%%%%%%
% PS: Please don't mind here until I'm done.

\section{Related Work and State of the Art}
\label{sec:relatedWork}

\subsection{GPU failure modes and what telemetry can and cannot show}
GPU failures in production environments span a wide spectrum, from gradual degradation (e.g., thermal throttling, power instability, clock oscillations) to abrupt device unavailability at the driver or interconnect layer. A practically important class of abrupt events is captured in NVIDIA's Xid error taxonomy, where specific Xid codes denote GPU channel resets and complete loss of device accessibility, including the commonly reported ``GPU has fallen off the bus'' condition (Xid 79). Such events are operationally relevant because they can occur with nominal numeric values immediately before the fault, after which device-level counters vanish or exporters stop producing per-GPU metrics \cite{nvidiaXidGuide}. Beyond detachments, large-scale reliability studies report that modern GPU faults often concentrate in specific subsystems (e.g., hardware components, interconnect, memory) and can propagate to node unavailability or job failure through diverse mechanisms, motivating multi-signal monitoring that does not assume a universal numeric precursor \cite{cui2025gpuResilience}. Complementary work on GPU memory faults shows that corruption and fault manifestation can be architecture- and workload-dependent and may not align cleanly with threshold-based telemetry assumptions, further supporting early-warning methods that integrate structural and cross-layer indicators rather than relying on value deviation alone \cite{sullivan2021micro}.

At the same time, reliability studies and fault-injection work emphasize that GPU failure behavior is strongly architecture-, workload-, and software-stack-dependent. Even when a system is instrumented, some pre-fault regimes may not yield stable numeric precursors in the exported metrics. This observation supports evaluation setups that do not assume deterministic ``drift before failure'' across all incidents, and instead separate drift-dominated regimes from detachment-class regimes in both methodology and reporting. \cite{gpuResilience2025}

\subsection{Telemetry pipelines for GPU nodes: DCGM and exporter semantics}
Monitoring based on DCGM (Data Center GPU Manager) provides a standardized interface for collecting GPU device telemetry at scale, including temperatures, power, utilization, clocks, and memory state. However, the semantics of missing DCGM-derived time series differ fundamentally from classical missing-at-random assumptions common in statistical monitoring. Missingness may arise from exporter behavior, transient driver resets, partial device enumeration loss, or failures in the communication path between the driver, DCGM, and the export layer, rather than from benign sampling gaps. NVIDIA's DCGM documentation explicitly delineates these component boundaries and failure modes, emphasizing that metric availability depends on the current driver and device state, not solely on exporter health \cite{nvidiaDcgmGuide}.

Empirical studies and operational experience further indicate that DCGM exporters may continue to emit node-level or subset GPU metrics even when one or more devices become inaccessible, leading to partial metric-family loss that can be easily misinterpreted as data-quality noise rather than a failure signal. This behavior complicates value-centric anomaly detection, as numeric telemetry may remain nominal for surviving devices while affected GPUs silently disappear from the metric stream. Similar challenges have been reported in large-scale GPU monitoring deployments, where exporter-level masking and delayed device re-enumeration obscure the temporal alignment between hardware failure and observable telemetry \cite{exadataZenodo}. These characteristics motivate treating metric disappearance, partial exporter output, and changes in per-scrape metric cardinality as first-class observability signals rather than preprocessing artifacts.

\subsection{Observability degradation and scrape-level indicators in Prometheus}
Prometheus exporters often expose scrape-level indicators (e.g., \texttt{up}) that describe the behavior of the monitoring pipeline itself as a measurable stochastic process rather than a passive transport layer. These indicators enable explicit quantification of monitoring quality degradation, including increased scrape latency, reduced sample payload size, intermittent scrape failures, and complete target unavailability. Prometheus exporters using Prometheus' exporter libraries create such time series with well-defined semantics out of the box, making them suitable for reproducible analysis and feature construction across deployments \cite{prometheusJobsInstances}.

Recent studies on large-scale monitoring systems emphasize that scrape-level behavior often degrades prior to hard failure, particularly under resource contention, exporter overload, network instability, or partial subsystem failure. In such cases, monitoring artifacts such as elevated scrape duration or reduced sample cardinality are not merely noise but reflect underlying system stress or loss of observability \cite{sigopsMonitoring2020}. This effect is amplified in GPU-heavy nodes, where exporters may selectively drop device-level metrics while remaining responsive at the HTTP level, yielding structurally valid but semantically degraded scrapes.

From an anomaly detection perspective, treating scrape degradation and metric disappearance as first-class signals departs from conventional preprocessing pipelines that discard incomplete samples or aggressively impute missing values. Prior work on telemetry-driven fault detection in distributed systems has shown that ignoring observability degradation can systematically delay detection of failure modes that do not manifest as value excursions in application or hardware metrics \cite{lou2022monitoringAnomalies}. In GPU detachment scenarios, where numeric telemetry often remains nominal until abrupt loss, scrape-level indicators may provide the only actionable early signal.

These observations motivate the inclusion of monitoring pipeline indicators as an explicit feature plane rather than auxiliary metadata. By modeling scrape behavior jointly with GPU and node telemetry, observability degradation becomes an analyzable failure symptom rather than a confounding artifact, enabling early-warning detection for failure regimes that are invisible to value-centric telemetry alone.

\subsection{Public GPU telemetry datasets and limitations for early-warning validation}
Large-scale public telemetry datasets have significantly accelerated research on GPU performance characterization, workload behavior, and anomaly detection. Datasets such as ExaData from CINECA’s Marconi100 provide curated, high-resolution GPU telemetry at scale, enabling detailed analysis of utilization, power, thermal behavior, and long-term trends across production-class systems \cite{exadataZenodo}. These resources have proven valuable for studying performance variability, efficiency, and hardware behavior under diverse workloads.

However, early-warning validation for node instability requires more than raw device-level time series. In particular, it requires access to operational context, including scheduler-reported state transitions (e.g., node drain, node down, failure reason codes), recovery actions, and maintenance events. Without such context, it is difficult to determine whether detected anomalies represent genuine pre-fault behavior, benign workload variation, or post-failure artifacts. This limitation is especially pronounced for abrupt failure modes, such as GPU detachments or interconnect-level faults, where numeric telemetry may remain nominal until failure.

Several studies on HPC reliability and anomaly detection emphasize that failure interpretation depends critically on correlating telemetry with system state and operational events. Datasets lacking scheduler-aligned failure context restrict evaluation to unsupervised anomaly discovery, without the ability to assess lead time, false-alarm relevance, or operational impact \cite{ferreira2019faultDiagnosis,maricq2021hpcAnomalies}. As a result, conclusions drawn from such datasets may overestimate predictive capability or conflate anomaly detection with post-hoc failure identification.

\subsection{Unlabeled anomaly detection and budgeted evaluation in HPC telemetry}
HPC anomaly detection frequently operates under incomplete labeling, non-stationary workloads, and heterogeneous nodes. In such settings, unsupervised detectors such as Isolation Forest and One-Class SVM are commonly used baselines because they can operate without explicit failure labels and provide continuous anomaly scores. However, operational deployment requires principled thresholding to avoid ad-hoc tuning. Budgeted alerting, where a fixed alert rate (e.g., 1\%) is enforced across methods, is a practical approach that makes detectors comparable under a shared triage budget. Within this constraint, the key question becomes whether additional feature planes (including pipeline degradation and structural disappearance) can increase lead time or improve detection coverage for failure regimes that are invisible to value-centric GPU telemetry alone.

%%%%%%%%%%                    %%%%%%%%%%

\section{Problem Statement and Hypothesis}
We study emerging GPU-node instability as a pre-fault condition that can precede hard failures and conventional alarms. In operational environments, such instability does not manifest uniformly. Empirical evidence from production telemetry indicates the presence of at least two distinct but operationally relevant regimes.

\textbf{Drift-dominated instability} refers to cases where GPU-node degradation develops gradually and produces weak but persistent numeric signals, such as thermal drift, power inefficiency, clock oscillations, or utilization-aware temperature trends. These signals may precede explicit failure or node state transitions and can be exploited for early warning under budgeted alerting.

\textbf{Detachment-class instability} refers to cases where GPU devices abruptly become unavailable at the driver or interconnect level (e.g., ``fallen off bus'' events). In these cases, conventional GPU telemetry often remains nominal until failure, and the dominant observable signal is structural: partial or complete loss of device-level metrics, reduced scrape payload size, increased scrape failures, or gaps in time series data.

The core hypothesis of this work is therefore twofold:

\begin{itemize}
    \item \textbf{H1:} Emerging GPU-node instability in drift-dominated regimes is observable through utilization-aware thermal-drift and trend signatures derived from GPU telemetry.
    \item \textbf{H2:} Emerging GPU-node instability in detachment-class regimes is observable primarily through observability-aware indicators that capture monitoring pipeline degradation and structural telemetry anomalies.
\end{itemize}

We further hypothesize that a joint modeling approach that integrates GPU-level signals with monitoring-pipeline indicators improves early warning capability across both regimes compared to GPU-only telemetry. 
A key practical constraint is the absence of explicit component-level failure labels in most operational telemetry streams.

Many public GPU telemetry datasets do not expose scheduler-level state transitions (e.g., node drain, node down, or failure reasons) that can be temporally correlated with GPU, power, and system-level metrics. Without such correlation, it is difficult to validate whether detected anomalies precede operationally meaningful events or simply reflect workload variation.

Therefore, we focus on telemetry collected from production GPU nodes at GWDG, where GPU, node, monitoring, and scheduler logs can be jointly analyzed. This enables construction of weak event proxies and supports investigation of both gradual and abrupt pre-fault behavior under realistic operational conditions.

\section{Dataset}

\subsection{Telemetry from GWDG GPU Nodes}
We use telemetry collected from production GPU nodes operated by the Gesellschaft für wissenschaftliche Datenverarbeitung mbH Göttingen (GWDG). The sanitized dataset used in this study has been publicly released on Zenodo \cite{gwdgGpuZenodo}. In contrast to many public telemetry corpora, this dataset enables direct correlation between GPU-level measurements, node-level signals, monitoring pipeline behavior, and scheduler-reported node state transitions.

The dataset includes:
\begin{itemize}
  \item GPU-level metrics exported via NVIDIA's DCGM exporter (e.g., temperature, power, utilization, clocks, memory state) \cite{nvidiaDcgmExporter},
  \item OS and node-level telemetry (e.g., CPU load, memory usage, system time) via the Prometheus Node Exporter included in Grafana Alloy as \texttt{prometheus.exporter.unix},
  \item Prometheus monitoring-pipeline indicators (e.g., scrape duration, scrape success, and per-scrape sample counts),
  \item Slurm metrics indicating node state transitions (e.g., \texttt{ALLOC}, \texttt{IDLE}, \texttt{DRAIN}, \texttt{DOWN}) and failure-related events gotten from a Prometheus Slurm Exporter (\url{https://github.com/vpenso/prometheus-slurm-exporter}) that has been patched to still work on recent versions of Slurm.
\end{itemize}

Slurm is configured to run the LBNL Node Health Check (NHC) (\url{https://github.com/mej/nhc}) every \SI{30}{min} to check nodes for failures not caught by jobs crashing and getting stuck in the epilog with ``Kill task failed''.
The NHC script, which determines which tests are done, checks the number of CPU cores, amount of RAM, network interfaces, filesystem mounts, etc. as well as GPUs.
The NHC script is how most GPU failures are caught.
Any problems found by NHC result in Slurm draining the node which shows up as a node state transition which is collected by the Prometheus Slurm Exporter.
The following GPU checks are done:

\begin{itemize}
    \item The \texttt{/dev/nvidia*} files are present for each GPU
    \item Run \texttt{nvidia-smi -L} to list all GPUs (``GPU fallen off bus'' often results in a timeout)
    \item Run \texttt{nvidia-smi --query-gpu fan.speed} to check that the fan speed is never ``Unknown Error'' (should be ``N/A'' on the water cooled GPUs)
    \item Run a small program that tests CUDA by checking that \texttt{cudaGetDeviceCount} is successful and that \texttt{cudaSetDevice} returns success or a memory allocation error (GPU has no VRAM left due to a job using all of it) on every GPU, otherwise returning failure. Note that the GWDG HPC admin team cannot recall a time this program caught a GPU failure that \texttt{nvidia-smi} didn't also catch, so this check appears to be redundant.
\end{itemize}

Beyond gradual performance and thermal effects, the GWDG dataset contains multiple instances of abrupt GPU-device detachments, commonly reported at the system level as ``GPU fallen off bus'' events. In these incidents, conventional GPU telemetry (temperature, power, utilization, clocks) often remains nominal prior to the event, while the dominant observable change is structural: partial or complete disappearance of GPU metrics, abrupt reductions in scrape sample counts, and increased monitoring gaps. These events are frequently node-local and may recur on the same physical host over extended time periods.

\paragraph{Quantitative summary of GPU-class and detachment incidents (forensic run).}
To characterize GPU-related instability, we executed a reproducible failure-forensic pass over the GWDG telemetry using a broad GPU failure-class filter \verb|failureClassRegex=^gpu|.
This pass matched 69 incident records across multiple GPU-related categories.
Of these, 15 incidents have had complete aligned telemetry archives obtained so far and were successfully processed.

Within this GPU-class set, we focus the present paper on the detachment subset labeled \texttt{gpu error / fallen off bus}.
This subset contains 7 incidents across three nodes (\texttt{ggpu142}, \texttt{ggpu149}, \texttt{cg1101}), of which 5 have complete telemetry and were processed, while 2 (both on \texttt{cg1101}) were excluded due to missing tidy telemetry.

Across all processed detachment incidents, alignment time $t_0$ was consistently derived from Prometheus scrape payload collapse (\texttt{scrapeCountDrop}) using a scrape interval of \SI{600}{s} and a dropout threshold of \SI{3000}{s}.
In every processed detachment case, GPU device-level telemetry disappeared partially or completely at or immediately after $t_0$, confirming observability collapse as the dominant observable manifestation of GPU detachment.

The presence of both broad GPU-class failures and a clearly defined detachment subset, together with detailed monitoring-pipeline telemetry and scheduler state transitions, makes the GWDG dataset particularly well-suited for studying observability-aware early warning under realistic operational conditions.

Table~\ref{tab:failureModes} summarizes the qualitative failure patterns observed in the GWDG production telemetry, highlighting the relationship between the presence or absence of numeric precursors and the dominant observable signals associated with each failure pattern.

\begin{table*}[t]
\centering
\caption{Qualitative characterization of GPU-node instability patterns observed in GWDG production telemetry.}
\label{tab:failureModes}
\setlength{\tabcolsep}{4pt}
\renewcommand{\arraystretch}{1.05}
\begin{tabular}{@{}lll@{}}
\toprule
Failure pattern & Numeric precursor & Dominant observable signal \\
\midrule
Thermal / efficiency drift &
Gradual (weak) &
Temperature drift and trend anomalies \\

Load-triggered instability &
Moderate (workload-correlated) &
Thermal and power excursions under load \\

GPU detachment (``fallen off bus'') &
None or negligible &
Loss of device metrics, scrape sample drop, gaps \\

Chronic detachment recurrence &
None per event &
Repeated structural anomalies on same node \\
\bottomrule
\end{tabular}
\end{table*}

\paragraph{Detachment-class incident selection.}
From the broader GPU-class forensic pass (\verb|^gpu|), we isolate the detachment subset labeled \texttt{gpu error / fallen off bus} in the operator incident catalog.
This subset contains 7 incidents spanning nodes \texttt{ggpu142}, \texttt{ggpu149}, and \texttt{cg1101}.
Five incidents (two on \texttt{ggpu142} and three on \texttt{ggpu149}) were successfully processed with complete telemetry archives.
Two incidents on \texttt{cg1101} could not be processed due to missing tidy telemetry files.

\subsection{Incident Catalog and Windowing}
\label{sec:incidentCatalog}

In addition to continuous telemetry, we use an operator-curated incident catalog to anchor analysis windows around operationally relevant events.
The failures in the catalog originate from real operational monitoring and scheduler-side evidence, including Slurm node-state transitions (node health script checks GPUs) and system-level logs associated with GPU-related failures. The incident catalog therefore provides coarse failure annotations rather than precise machine-aligned labels.

Each incident entry specifies the affected node, incident date which is either when it happened or when it was noticed (e.g. failure on a Saturday not noticed till Monday), free-text description, failure category, and asymmetric collection bounds (\texttt{beforeHours}, \texttt{afterHours}), which define the raw telemetry interval (\texttt{collectStart}, \texttt{collectEnd}) used for analysis.
Since the incident date in the catalog could be on the day of the error or when it was noticed, pre-processing had to be done to find when the incident took place more precisely to determine the \texttt{collectStart} and \texttt{collectEnd} to use.
For each incident entry in the catalogue,

\begin{enumerate}
    \item The times of all slurm node-state transitions on the from OK (idle, alloc, mix) to failure (drain, draining, down, no response, rebooting) were found. The failures either come from the node health script check which checks the GPUs empirically (every \SI{30}{min}), a GPU job failing, or the node no longer being responsive.
    \item If there is at least one transition from OK to failure on the day logged in the incident catalog, the first one is chosen as the incident time.
    \item If there is at least one transition from OK to failure in the 3 days prior to the day logged in the incident catalog, the last one is chosen as the incident time.
    \item Otherwise the incident is discarded due to not being able to identify when the incident occurred (most likely this is from not logging the incident in the catalog till more than 3 days after it occurred).
\end{enumerate}

The catalog spans January 2025 to early February 2026 and covers a diverse set of node-level failure modes, including kernel panics and softlocks, node hangs and resets, watchdog-triggered failures, network and interconnect degradation, memory and ECC-related faults, and GPU detachment events.
This diversity enables evaluation of observability-aware methods beyond GPU-specific failures, even though this paper focuses its forensic case studies on GPU detachment.

Table~\ref{tab:incidentCategories} summarizes the category distribution in the incident catalog used in this study.

% \begin{table}[t]
% \centering
% \caption{Incident categories and counts in the analyzed GWDG incident catalog (Jan 2025--Jan 2026).}
% \label{tab:incidentCategories}
% \setlength{\tabcolsep}{5pt}
% \renewcommand{\arraystretch}{1.05}
% \begin{tabular}{@{}lr@{}}
% \toprule
% Incident category & Count \\
% \midrule
% GPU error / fallen off bus & 7 \\
% Node hang / offline / reset & 6 \\
% Kernel panic / softlock & 2 \\
% Watchdog / service-kill & 2 \\
% Network / IB / OPA & 3 \\
% Memory / ECC / MCE & 1 \\
% Other & 3 \\
% \midrule
% Total & 24 \\
% \bottomrule
% \end{tabular}
% \end{table}

\begin{table}[t]
\centering
\caption{GPU-related incident categories and counts in the analyzed GWDG incident catalog.}
\label{tab:incidentCategories}
\setlength{\tabcolsep}{5pt}
\renewcommand{\arraystretch}{1.05}
\begin{tabular}{@{}lr@{}}
\toprule
Incident category & Count \\
\midrule
GPU error / problem & 31 \\
GPU fell off bus & 24 \\
GPU unknown & 5 \\
GPU lost & 3 \\
GPU ECC & 2 \\
GPU failed & 2 \\
GPU timeout & 1 \\
GPU handle error & 1 \\
\midrule
Total & 69 \\
\bottomrule
\end{tabular}
\end{table}

\subsection{Why Not Marconi100 ExaData}
Public datasets such as CINECA’s Marconi100 ExaData provide large-scale GPU telemetry and have been instrumental in advancing research on GPU performance and reliability. However, in the context of early-warning detection for node instability, ExaData exhibits two critical limitations:

\begin{itemize}
  \item \textbf{Lack of scheduler-correlated failure context:} ExaData does not expose node state timeseries according to the job scheduler indicating node drain, node down, or failure reasons that can be temporally aligned with GPU, power, and system metrics.
  \item \textbf{Limited ability to validate pre-fault behavior:} Without such correlated operational state transitions, it is not possible to determine whether detected anomalies precede meaningful node-level events or simply reflect workload variation.
\end{itemize}

Because our goal is to study pre-fault instability and early warning, we require datasets where telemetry can be correlated with node state evolution. The GWDG dataset satisfies this requirement and is therefore used for all baseline experiments in this paper.

\subsection{Reproducible Slice Definition}
To satisfy reproducibility expectations, each experiment is defined by an explicit slice specification:
\begin{itemize}
  \item Node identifiers included in the slice.
  \item Time coverage: start and end timestamps after filtering.
  \item Native sampling interval: median timestamp delta prior to windowing.
  \item Windowing: window length $w$ and stride $s$ used to form aggregated windows.
  \item Per-node sampling cap and random seed.
\end{itemize}

\paragraph{Reproducibility summary.}
% Baseline detectors use fixed window aggregation with specific length $w$ and stride $s$ as configured in the baseline run (exported with the artifact metadata).
% After filtering, the median native sampling interval is approximately \SI{600}{s} (timestamps in the cleaned per-node archives end at \texttt{xx:50}, consistent with a 10-minute cadence).
% The evaluated slice spans 7 GPU nodes (\texttt{ggpu121}, \texttt{ggpu129}, \texttt{ggpu139}, \texttt{ggpu142}, \texttt{ggpu143}, \texttt{ggpu144}, \texttt{ggpu149}) and covers \texttt{2025-02-03 00:00:00} to \texttt{2026-01-21 23:50:00} based on \texttt{minTime--maxTime} in the ETL manifest (approximately 353 days).

Baseline detectors use fixed window aggregation with specific length $w$ and stride $s$ as configured in the baseline run (exported with the artifact metadata).
After filtering, the median native sampling interval is approximately \SI{600}{s}, consistent with the Prometheus scrape interval (timestamps in the cleaned per-node archives typically end at \texttt{xx:50}, reflecting a 10-minute cadence).
The evaluated slice spans 7 GPU nodes (\texttt{ggpu121}, \texttt{ggpu129}, \texttt{ggpu139}, \texttt{ggpu142}, \texttt{ggpu143}, \texttt{ggpu144}, \texttt{ggpu149}) and covers \texttt{2025-02-03 00:00:00} to \texttt{2026-01-21 23:50:00} according to the ETL manifest (\texttt{minTime--maxTime}), corresponding to approximately 353 days of observations.
Per-node GPU inventory indicates 4 GPUs per node in this slice (28 GPUs total).
Isolation Forest and One-Class SVM hyperparameters are exported alongside the evaluation outputs and included in the artifact metadata.

\subsection{Per-Node Sampling and Representativeness}
To control computational cost while preserving cross-node diversity, the baseline experiment uses per-node sampling with a fixed cap of 500 windows per node, producing 8000 merged windows. This prevents a small number of high-volume nodes from dominating the slice. Representativeness is evaluated by reporting per-node coverage and sensitivity to random seed in the final metadata export.

\subsection{Missingness and Gap Statistics}
Telemetry incompleteness is a first-order property of operational monitoring. Missingness ratios and gap-length distributions are computed directly from each slice and reported alongside experimental results.

\section{Methodology}
\subsection{Windowing and Notation}
Raw time series are aligned and aggregated into fixed windows of length $w$ with stride $s$. For each node $n$ and GPU $g$, the window-level feature vector is:
\[
\mathbf{x}_{n,g}(t) = [\mathbf{x}^{gpu}_{n,g}(t), \mathbf{x}^{pipe}_{n}(t), \mathbf{x}^{os}_{n}(t)].
\]
Here, $\mathbf{x}^{gpu}_{n,g}(t)$ is GPU-local telemetry for GPU $g$ on node $n$, $\mathbf{x}^{pipe}_{n}(t)$ is node-level monitoring-pipeline (observability) features shared by all GPUs on node $n$, and $\mathbf{x}^{os}_{n}(t)$ denotes node-level OS telemetry.
We define $t_0$ as the incident time.

All methods consume the same windowed feature matrix for a given slice specification.

\paragraph{Baseline windowing configuration.}
In the baseline experiments, raw telemetry is aggregated using a fixed window length of $w=\SI{60}{min}$ with stride $s=\SI{10}{min}$.
After filtering and alignment, the median native sampling interval is \SI{600}{s}, corresponding to 10 times the Alloy scrape interval (\SI{60}{s}).
Lead time is therefore reported in windows, where one window corresponds to \SI{10}{min}.

\paragraph{Time-scale separation.}
We distinguish three time scales used throughout the analysis:
(i) \emph{incident collection windows} defined by the operator catalog after the proprocessing stop to more precisely find when the incident occurred (\texttt{beforeHours}/\texttt{afterHours}, e.g., 24\,h/2\,h), which determine the raw telemetry interval extracted around each incident;
(ii) \emph{detector feature windows} with length $w$ and stride $s$, used to aggregate time series for anomaly detection;
and (iii) a compact \emph{forensic comparison window} consisting of a \SI{30}{min} baseline interval and a \SI{5}{min} interval adjacent to $t_0$ (as configured by \texttt{tAfterMin}), used to rank metric shifts and structural disappearance around detachment events.
This separation avoids conflating data collection bounds with detector aggregation and forensic comparison.

\subsection{Preprocessing and Alignment}
We apply the following deterministic preprocessing steps:
\begin{itemize}
  \item Timestamp normalization and alignment to a common timeline per node.
  \item Aggregation per window using mean, standard deviation, minimum, maximum, and slope.
  \item Robust scaling for learning-based models using per-feature median and MAD.
  \item Missing-value handling with explicit missingness reporting when available and conservative imputation when required.
\end{itemize}

\subsection{GPU Telemetry Feature Plane}
We extract and aggregate (mean, max, min, std, slope) from GPU-related metrics. In the baseline run, GPU-plane features are primarily derived from memory temperature channels and their temporal structure, including drift and slope components (Section~\ref{sec:signature}).

\subsection{Monitoring Pipeline Feature Plane}
We quantify monitoring pipeline behavior using scrape- and probe-level indicators when available in the selected archive subset(s). Unlike conventional monitoring practice where missing or incomplete telemetry is often treated purely as a data quality issue, we explicitly model monitoring degradation as a first-class anomaly signal.

The monitoring pipeline feature plane captures indicators such as scrape duration, scrape success, and per-scrape sample counts. Abrupt reductions in scraped sample volume, increased scrape latency, intermittent scrape failures, and persistent gaps in time series data are treated as structural anomalies that may indicate underlying subsystem failure rather than benign monitoring noise.

This design is motivated by detachment-class GPU failures observed in production telemetry, where device-level metrics disappear while exporters and nodes may otherwise remain responsive. In such cases, monitoring degradation is often the dominant and sometimes the only observable manifestation of the failure. Pipeline-plane evaluation is therefore reported only on slices where these metrics exist and are non-empty after feature construction, ensuring valid cross-plane comparisons.

\subsection{Instability Signatures and Robustness Constraints}
\label{sec:signature}
Divergence ratios can be unstable when utilization is low and when temperature lags workload. In the current baseline, the instability signature is intentionally constrained to robust thermal-drift and ambient-drift indicators that remain available and stable under the sampled slice conditions.
Power--utilization divergence features are part of the full framework but are not included in the current baseline signature.

\subsubsection{Signature feature set}
The baseline signature uses 16 columns:
\begin{itemize}
  \item GPU memory temperature drift features: per-GPU average, minimum, and maximum drift for four GPUs (12 total).
  \item Ambient drift features: average, minimum, and maximum drift (3 total).
  \item A rolling-slope feature capturing sustained memory temperature trend (\texttt{memTemp\_rollSlope\_32}) (1 total).
\end{itemize}

\subsection{Detectors and Fixed Settings}
The baseline compares three detectors under a fixed alert budget of 1\%:
\begin{itemize}
  \item Robust z-score scoring.
  \item Isolation Forest.
  \item One-Class SVM.
\end{itemize}
Smoothing is applied as a rolling mean with window size 5, and lead time is evaluated with a fixed lookback of 48 windows.

\section{Evaluation}
\label{sec:evaluation}

\subsection{Alert Budget and Thresholding}
All detectors produce a continuous anomaly score per window. Alerts are generated by choosing a threshold such that only the top 1\% of scores trigger alerts. Budgeted alerting removes ad-hoc threshold tuning and ensures comparable alert volume across detectors.

\subsection{Weak-Event Construction for Unlabeled Evaluation}

Because explicit component-level failure labels are unavailable in the analyzed telemetry, we define weak events as persistent excursions of the GPU-derived instability signature. 
Formally, a weak event is a contiguous run of windows in which the signature exceeds the 0.99 quantile threshold for at least three consecutive windows.

This proxy is intended to capture drift-dominated instability regimes with gradual numeric deviation before operational impact. It is not intended to represent detachment-class failures, where the dominant pre-fault manifestation is structural observability degradation rather than a stable excursion in conventional GPU telemetry.

The study intentionally focuses on a small number of fully aligned detachment incidents and evaluates detectors under unlabeled operational telemetry using weak-event proxies, reflecting the empirical property that many GPU detachment failures provide little or no stable numeric precursor and therefore constrain classical predictive evaluation.

% Explicit component-level failure labels are unavailable in the analyzed telemetry. 
% We therefore construct weak events as persistent excursions of the instability signature derived from GPU telemetry. 
% A weak event is defined as a contiguous run of windows where the signature exceeds the 0.99 quantile threshold for at least three consecutive windows.

% The study intentionally focuses on a small number of fully aligned detachment incidents and evaluates detectors under unlabeled operational telemetry using weak-event proxies, reflecting the empirical property that many GPU detachment failures provide little or no stable numeric precursor and therefore constrain classical predictive evaluation.

% This definition captures drift-dominated instability patterns in which gradual thermal or efficiency anomalies precede operational impact. By design, it does not cover detachment-class failures where GPU devices abruptly become unavailable and conventional telemetry remains nominal until disappearance. In such cases, the dominant observable signal is structural (e.g., loss of device metrics or scrape payload collapse) rather than numeric excursion.

\subsection{Empirical Weak-Event Characteristics}
To quantify pre-fault feature behavior in the weak-event slice, we summarize incident-anchored feature shifts extracted from \texttt{weak\_events\_summary.csv}. Each row corresponds to a pre-failure window aligned to $t_0$, the operator incident time after the pre-processing search for the actual failure time. Table~\ref{tab:weakEventSummary} reports the number of available feature signals (\texttt{numSignalsLong}) and the dominant feature shifts ranked by absolute delta within the forensic comparison window.

\begin{table*}[t]
\centering
\caption{Weak-event summary extracted from \texttt{weak\_events\_summary.csv}. 
For each incident-anchored weak event (\texttt{label=pre\_failure}), we report the number of available signals (\texttt{numSignalsLong}) and the dominant delta-ranked feature shifts within the forensic comparison window. Variance-shift statistics (\texttt{diffStd}) are frequently zero in this slice and therefore do not provide a stable ranking axis.}
\label{tab:weakEventSummary}
\setlength{\tabcolsep}{4pt}
\renewcommand{\arraystretch}{1.05}
\scriptsize
\begin{tabular}{@{}llllp{7.2cm}@{}}
\toprule
Node & $t_0$ (UTC) & Category & \makecell{Signals\\(\texttt{numSignalsLong})} & Top signals by $|\Delta|$ \\
\midrule
ggpu121 & 2025-02-09 15:00 & gpu error/problem & 70 &
\texttt{node\_memory\_MemAvailable\_bytes} (\(\Delta=13615104\)), 
\texttt{node\_load1} (\(\Delta=-0.09\)), 
\texttt{node\_load5} (\(\Delta=-0.04\)), 
\texttt{node\_load15} (\(\Delta=-0.03\)) \\
\midrule
ggpu139 & 2025-03-21 09:45 & gpu fell off bus & 510 &
\texttt{nodes\_total\_gpus\_when\_good} (\(\Delta=-4\)) \\
\midrule
ggpu142 & 2025-02-16 12:50 & gpu fell off bus & 340 &
\texttt{node\_memory\_MemAvailable\_bytes} (\(\Delta=-3244032\)) \\
\midrule
ggpu149 & 2025-06-12 07:30 & gpu fell off bus & 350 &
\texttt{node\_memory\_MemAvailable\_bytes} (\(\Delta=111666880512\)), 
\texttt{node\_load15} (\(\Delta=-0.05\)), 
\texttt{node\_load5} (\(\Delta=-0.01\)) \\
\bottomrule
\end{tabular}
\end{table*}

Across the weak-event slice, variance-shift statistics are frequently zero and do not provide stable ranking. Observable pre-fault differences are therefore dominated by delta-level feature shifts and structural feature disappearance rather than sustained numeric drift. This empirical property of the slice explains why drift-based weak-event lead time is limited in the baseline results.
Additionally, the dominant delta-level features are often just changes in the host CPU load or RAM usage, rather than GPU metrics.
These are potentially from a job allocating a lot of memory right before failure or memory deallocation right around the failure, possibly due to the job failing due to the GPU failure and the processes dying.

\subsection{Incident-Anchored Evaluation for Detachment Failures}
Because detachment-class failures are not captured by signature-derived weak events, we additionally perform incident-anchored evaluation using the operator incident catalog after pre-processing to find the approximate incident time. For these cases, the reference time $t_0$ is aligned via scrape payload collapse (\texttt{scrapeCountDrop}). Detectors are evaluated on whether observability-aware indicators trigger within the incident window before operational impact is recorded.

\begin{table*}[t]
\caption{Incident-anchored pre-fault observability behavior for GPU detachment incidents.
Dominant signals are structural (GPU telemetry disappearance) rather than stable numeric precursors.}
\label{tab:detachmentObservabilitySummary}
\centering
\begin{tabular}{@{}llll@{}}
\toprule
Node & Alignment time $t_0$ & Category & Dominant pre-fault observability signal \\
\midrule
ggpu142 & 2025-02-16 12:50 & GPU detachment &
Partial GPU metric-family loss and scrape payload collapse \\
ggpu142 & 2025-03-21 09:10 & GPU detachment &
GPU telemetry disappearance with preceding temperature/utilization activity \\
ggpu149 & 2025-03-21 10:40 & GPU detachment &
Abrupt GPU metric disappearance with scrape degradation \\
ggpu149 & 2025-06-12 07:30 & GPU detachment &
GPU telemetry loss preceding delayed operator detection \\
ggpu149 & 2026-01-18 12:40 & GPU detachment &
GPU metric disappearance without stable numeric precursor \\
\bottomrule
\end{tabular}
\end{table*}

Table~\ref{tab:detachmentObservabilitySummary} confirms that detachment-class failures exhibit little or no stable numeric precursor in GPU telemetry. Instead, the dominant observable signal is structural, consisting of GPU metric disappearance, scrape payload reduction, and monitoring gaps around $t_0$. This behavior is consistent across all processed detachment incidents.

\begin{table*}[t]
\centering
\caption{Detachment-class incidents used for forensic alignment (\texttt{failureClass}=\texttt{gpu error / fallen off bus}). The alignment time $t_0^{used}$ is derived from \texttt{scrapeCountDrop}.}
\label{tab:detachmentIncidentsForensics}
\setlength{\tabcolsep}{4pt}
\renewcommand{\arraystretch}{1.05}
\begin{tabular}{@{}llll@{}}
\toprule
Node & $t_0^{incident}$ & $t_0^{used}$ & Tidy archive \\
\midrule
ggpu142 & 2025-02-17 00:00 & 2025-02-16 12:50 & ggpu142\_2025-02-17\_gpus-fallen-off-bus\_tidy.csv.bz2 \\
ggpu142 & 2025-03-21 00:00 & 2025-03-21 09:10 & ggpu142\_2025-03-19\_gpu-problem\_tidy.csv.bz2 \\
ggpu149 & 2025-03-21 00:00 & 2025-03-21 10:40 & ggpu149\_2025-03-21\_gpus-dropped-off-bus\_tidy.csv.bz2 \\
ggpu149 & 2025-06-12 00:00 & 2025-06-12 07:30 & ggpu149\_2025-06-12\_gpus-dropped-off-bus\_tidy.csv.bz2 \\
ggpu149 & 2026-01-19 00:00 & 2026-01-18 12:40 & ggpu149\_2026-01-19\_gpus-dropped-off-bus\_tidy.csv.bz2 \\
\bottomrule
\end{tabular}
\end{table*}

Two additional detachment incidents on \texttt{cg1101} are omitted because their tidy telemetry archives were missing and forensic alignment could not be executed.

This dual evaluation strategy reflects the presence of two empirically distinct failure regimes: gradual drift with numeric precursors and abrupt detachment dominated by structural observability anomalies.

\subsection{Lead Time Definition}
For each weak event, lead time is defined as the number of windows between the first alert and the start of the weak event, subject to a lookback horizon of 48 windows. We report average, median, and maximum lead points. In this baseline, lead time is expressed in windows; conversion to minutes depends on the native sampling interval and window stride.

\section{Results and Case Studies}
\paragraph{Case-study selection}
We report detachment-class case studies from nodes \texttt{ggpu142} and \texttt{ggpu149}, spanning February 2025 to January 2026, as identified in the incident catalog in Table~\ref{tab:detachmentIncidentsForensics}.
Notably, for \texttt{ggpu149} on 12 June 2025, the operator incident record indicates that NHC detected the condition many hours after the underlying GPU detachment event.
Forensic alignment shows that GPU telemetry disappearance and scrape degradation occurred near $t_0$, well before operator detection.
This gap illustrates the operational importance of observability-aware indicators that react to telemetry collapse earlier than delayed health checks.

\subsection{Baseline Results Under Fixed Alert Budget}
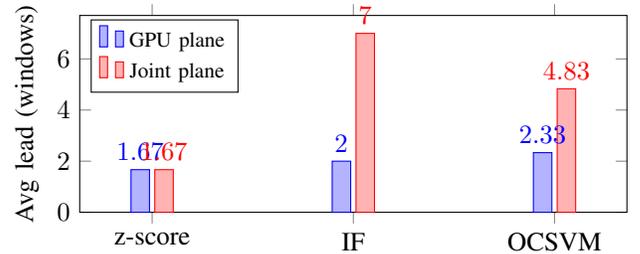
\begin{figure}[t]
\centering
\begin{tikzpicture}
\begin{axis}[
    ybar,
    bar width=7pt,
    width=\columnwidth,
    height=4.2cm,
    enlarge x limits=0.18,
    ylabel={Avg lead (windows)},
    symbolic x coords={z-score,IF,OCSVM},
    xtick=data,
    legend style={font=\footnotesize, at={(0.02,0.98)}, anchor=north west},
    nodes near coords,
    nodes near coords align={vertical},
    ymin=0
]
\addplot coordinates {(z-score,1.667) (IF,2.000) (OCSVM,2.333)};
\addplot coordinates {(z-score,1.667) (IF,7.000) (OCSVM,4.833)};
\legend{GPU plane,Joint plane}
\end{axis}
\end{tikzpicture}
\caption{Average lead time under a fixed 1\% alert budget. Joint (GPU + observability) increases early warning, especially for Isolation Forest (IF), compared to GPU-only features.}
\label{fig:leadAvg}
\end{figure}

Table~\ref{tab:resultsFilled} and figure~\ref{fig:leadAvg} summarize the baseline comparison using the metrics exported by the evaluation output. We report only fields that are present in the baseline results.

We compare robust z-score scoring, Isolation Forest (IF), and One-Class SVM (OCSVM).

\begin{table}[t]
\centering
\caption{Baseline plane comparison at 1\% alert budget. Lead time is reported in windows within a lookback of 48 windows.}
\label{tab:resultsFilled}
\setlength{\tabcolsep}{3pt}
\renewcommand{\arraystretch}{0.95}
\scriptsize
\begin{tabular}{@{}llrrrrr@{}}
\toprule
Plane & Method &
\makecell{Avg\\Lead} &
\makecell{Median\\Lead} &
\makecell{Max\\Lead} &
\makecell{Avg\\RunLen} &
Runs \\
\midrule
GPU   & z-score         & 1.667 & 0.0 & 8.0  & 4.706 & 17 \\
GPU   & Isolation Forest& 2.000 & 0.0 & 10.0 & 8.000 & 10 \\
GPU   & One-Class SVM   & 2.333 & 0.0 & 11.0 & 7.273 & 11 \\
\hline
Joint & z-score         & 1.667 & 1.0 & 5.0  & 5.000 & 16 \\
Joint & Isolation Forest& 7.000 & 1.5 & 29.0 & 4.706 & 17 \\
Joint & One-Class SVM   & 4.833 & 0.0 & 26.0 & 8.000 & 10 \\
\bottomrule
\end{tabular}
\end{table}

Table \ref{tab:resultsFilled} shows that using both GPU and observability features together (“Joint”) gives earlier warnings than using GPU data alone. In particular, Joint Isolation Forest achieves the highest average lead time (7 windows) and the longest maximum lead (29 windows). GPU-only detectors typically trigger close to or after event onset (median lead = 0), meaning they often detect instability late. Having this, combining observability signals with GPU telemetry improves early detection without increasing the alert budget.
The lead time for joint IF and OCSVM are potentially sufficient that if they could be extended to a live monitoring system, compute jobs on a node right before GPU failure could be signaled to warn them that GPU failure is imminent so that suitably designed jobs have a chance to take snapshots of their current progress which they could later resume in another job on a different node, potentially even submitting the follow-up job right from the job as it is cleaning up (not all workloads support snapshots, but many do).
Though getting users to design their jobs to support this might actually be the harder part.

\subsection{Engineering Interpretation of the Baseline}
The baseline yields three concrete observations under a fixed alert budget and a fixed lead lookback:
\begin{itemize}
  \item \textbf{Joint features increase early-warning depth for learning-based detectors:} the joint Isolation Forest achieves the highest average lead (7 windows) and a higher median lead (1.5 windows) than any GPU-only detector. This indicates that cross-plane aggregation can convert weak drift into earlier alerts without increasing the alert budget.
  \item \textbf{Median lead is often zero under weak events:} most configurations show a median lead of 0.0 windows, meaning that for at least half of the weak events the first alert occurs at or after event onset. This is expected under strict budgeted alerting and a conservative weak-event definition, and it highlights the need for a larger slice and additional plane coverage to separate early warning from event detection.
  \item \textbf{Alert episode structure differs by detector:} for example, joint Isolation Forest yields more fragmented alerting (17 runs with shorter average run length) than joint One-Class SVM (10 runs with longer average run length). This matters operationally because alert fragmentation increases triage overhead even when the alert rate is fixed.
\end{itemize}

\section{Discussion}
\label{sec:discussion}

\subsection{What is Reproducible Now and What Remains}
From the exported forensic outputs, we can reproduce and report:
\begin{itemize}
  \item GPU-class forensic scope: 69 matched incidents, 15 processed, 54 missing tidy archives
  \item Detachment subset: 7 matched incidents, 5 processed, 2 missing tidy archives
  \item alignment mechanism: $t_0$ derived from scrape payload collapse (\texttt{scrapeCountDrop})
  \item alert budget (0.01), smoothing window (5), and lead lookback (48 windows)
  \item weak-event extraction parameters (quantile 0.99, minRun 3)
  \item plane sizes through feature counts (GPU: 17 features, Joint: 81 features)
  \item budgeted evaluation metrics in Table~\ref{tab:resultsFilled}
\end{itemize}

What remains to be exported before we can populate the dataset-specification tables is:
\begin{itemize}
  \item distinct node count and GPU coverage in the slice
  \item explicit time range of the slice and native sampling interval
  \item window length $w$ and stride $s$ used for aggregation
  \item missingness and gap statistics per plane
  \item detector hyperparameters for learning-based models
\end{itemize}

\subsection{Dataset Choice and External Validity}
Our choice of using GWDG production telemetry rather than public datasets such as Marconi100 ExaData is driven by the need for cross-layer correlation between telemetry and scheduler-level node state derived from compute job failures and empirically checking the GPUs by active probing (\texttt{nvidia-smi} and allocating VRAM). While ExaData offers valuable large-scale GPU measurements, it does not expose Slurm (or equivalent) logs indicating node drain, node down, or failure-reason events that can be temporally aligned with GPU, power, and node-level metrics. This limits the ability to assess whether detected anomalies precede operationally meaningful node state transitions.

The proposed methodology is dataset-agnostic and compatible with ExaData-style telemetry. As more public datasets include scheduler and failure-context information, direct replication of our experiments on external corpora will become feasible.

\subsection{Failure Modes Without Numeric Precursors}
A central observation from the GWDG production telemetry is that a significant class of GPU failures exhibits little or no numeric precursor in conventional GPU metrics such as temperature, power consumption, clocks, or utilization. In these detachment-class failures, GPUs may operate within nominal ranges until they abruptly become unavailable at the driver or interconnect level.

From an anomaly detection perspective, the absence of detectable numeric drift prior to failure is not a shortcoming of the detection framework but a property of the underlying failure mode. Certain hardware faults, such as marginal PCIe connectivity, firmware state corruption, or power-delivery instability, may not produce measurable deviations in available telemetry before failure. In such cases, the expectation of gradual precursors is unrealistic, and value-based anomaly detection methods are fundamentally limited.

\subsection{Structural Anomalies as First-Class Signals}
The analyzed detachment incidents consistently demonstrate that structural telemetry anomalies are the dominant observable manifestation of failure.
Across all processed cases, GPU device-level metrics disappeared or degraded abruptly at alignment time $t_0$, while conventional numeric telemetry showed weak or inconsistent precursor behavior due to sparse sampling and normalization sensitivity. 
Slice-level variance-shift statistics were frequently zero in this regime, indicating that pre-fault differences are not expressed as stable volatility change but primarily as structural observability collapse with only weak delta-level feature shifts.
These signals originate at the observability layer and are frequently dismissed as monitoring artifacts or data quality issues in traditional workflows.

In sensitive infrastructures, however, observability degradation is tightly coupled to subsystem health. When devices become unreachable, exporters may continue running while silently dropping affected metric families, leading to partial observability collapse without explicit failure alarms. Treating metric disappearance and scrape degradation as first-class anomaly signals enables detection of failure modes that are otherwise invisible to value-centric monitoring approaches.

\subsection{Recurrence and Host-Level Hazard}
An additional operational insight from the incident catalog is the recurrence of detachment-class failures on specific physical nodes over extended time periods.
Node \texttt{ggpu142} experienced two GPU detachment incidents within approximately one month (February--March 2025), while \texttt{ggpu149} experienced three detachment incidents over a ten-month period (March 2025, June 2025, and January 2026).

From a reliability engineering perspective, recurrence is a more informative hazard signal than the severity of any single event. Nodes that repeatedly experience device detachments are unlikely to self-heal and pose an elevated risk to workload stability and operational efficiency. Consequently, anomaly detection systems for sensitive infrastructures should incorporate host-history aggregation and recurrence-aware scoring, enabling proactive intervention such as node quarantine, reallocating nodes to lower priority and shorter duration workloads that are easier to redo (e.g. short runs for testing codes before longer computations), hardware derating (e.g. reducing the maximum clock frequencies of some components so that they can still operate but at reduced performance if that reduces the chance of failure), hardware replacement, or retirement before catastrophic failure occurs.

%%%%%%%%%% RELATED WORKS STUFF %%%%%%%%%%

\subsection{Positioning of this work}
Prior work on GPU reliability and anomaly detection has largely focused on value-centric analysis of device telemetry, including temperature, power, utilization, and performance counters \cite{gpuResilience2025,exadataZenodo}.
These approaches are effective for drift-dominated regimes, where gradual degradation produces measurable numeric precursors.
However, they implicitly assume that failure-relevant signals remain observable and that deviations manifest as changes in metric values rather than in metric availability.

This work addresses a complementary and underexplored failure regime in which GPUs fail quietly at the driver or interconnect layer, producing little or no numeric precursor in conventional telemetry \cite{nvidiaXidGuide}.
In such cases, the dominant observable signal is structural, including partial or complete disappearance of device metrics, reduced scrape payload size, increased scrape latency, and monitoring gaps \cite{prometheusJobsInstances,nvidiaDcgmGuide}.
These effects are typically treated as monitoring artifacts or data quality issues and are therefore excluded from most anomaly detection pipelines.

We position this work as an observability-aware extension to existing GPU anomaly detection frameworks.
Rather than replacing value-centric methods, we integrate them with monitoring-pipeline indicators that explicitly model telemetry degradation as a first-class signal \cite{prometheusJobsInstances}.
This joint perspective enables early-warning detection across both drift-dominated and detachment-class failure regimes under a fixed alert budget and without relying on explicit component-level failure labels.

By grounding the analysis in production telemetry with incident-aligned scheduler context, this work bridges the gap between GPU-centric anomaly detection and operational failure forensics, providing a practical methodology for early-stage reliability assessment in large-scale GPU clusters.

%%%%%%%%%%                     %%%%%%%%%%

\section{Conclusion}
This paper investigated early-warning detection of GPU-node instability under realistic operational conditions where explicit component-level failure labels are unavailable and conventional threshold alarms frequently fail to trigger. Using production telemetry from GWDG GPU nodes, we demonstrated that quiet GPU failures manifest in at least two distinct regimes: drift-dominated instability, where weak thermal or efficiency-related signals emerge gradually, and detachment-class instability, where GPU devices abruptly become unavailable with little or no numeric precursor. The public sanitized dataset accompanying this study is available on Zenodo \cite{gwdgGpuZenodo}.
Our forensic analysis of detachment-class incidents shows that value-centric anomaly detection based solely on GPU telemetry is insufficient to capture a significant class of operational failures.
In all processed detachment cases, GPUs became unavailable abruptly and the earliest reliable signal was structural: disappearance of device-level metrics and collapse of scrape payload integrity.
Numeric precursors in conventional telemetry were weak, inconsistent, or absent, and variance-shift statistics were frequently zero in the analyzed slice, reflecting both the abrupt physical failure mode and the sparse sampling regime of operational monitoring.

Treating observability degradations as first-class anomaly signals is essential for detecting failures that are otherwise invisible to conventional monitoring approaches.

Within a reproducible and budgeted evaluation framework, we found that joint modeling of GPU-level signals and monitoring-pipeline indicators can increase early-warning lead time for learning-based detectors compared to GPU-only detection under the same alert budget. At the same time, the analysis revealed that many failures offer limited or no deterministic numeric precursor, underscoring the importance of honest evaluation practices that distinguish early warning from event detection and avoid overstating predictive capability. These results indicate that early warning in GPU infrastructures benefits from combining heterogeneous telemetry sources and from explicitly modeling missing or degraded observability signals, rather than relying solely on pattern deviations in available numeric metrics.

Beyond individual events, we observed that recurrence of detachment-class failures on the same physical nodes constitutes a strong hazard signal. This highlights the need for host-history aggregation and recurrence-aware scoring in anomaly detection systems for sensitive infrastructures, enabling proactive intervention such as node quarantine or hardware replacement before repeated failures degrade system reliability.

This work advances observability-aware anomaly detection by explicitly accounting for structural telemetry anomalies, monitoring pipeline degradation, and failure heterogeneity. While the experiments focus on GPU nodes, the proposed methodology generalizes to other hardware-intensive and safety-critical infrastructures where abrupt device detachments and observability loss are common. Future work will extend the framework to multi-archive slices, explicit structural weak-event definitions, and broader cross-system validation as richer failure-context datasets become available.

\section*{Acknowledgment}
This work was supported by the KISSKI Project funded by the German Federal Ministry of Education and Research (BMBF) under grant 01|S22093A (Foerderkennzeichen). We thank GWDG for access to production telemetry and operational expertise. We also thank Mr. Timon Vogt for his contribution in providing support for node data collection.

\bibliographystyle{IEEEtran}
\bibliography{references}

\end{document}